\documentclass[fleqn,10pt]{wlscirep}

\graphicspath{{./}{./images/}{./images_all/}{./images_all/multiferroic_read_write/}}

\title{Separating read and write units in multiferroic devices}

\author[1,*]{Kuntal Roy}
\affil[1]{School of Electrical and Computer Engineering, Purdue University, West Lafayette, Indiana 47907, USA}
\affil[*]{royk@purdue.edu}

\begin{abstract}
Strain-mediated multiferroic composites, i.e., piezoelectric-magnetostrictive heterostructures, hold profound promise for energy-efficient computing in beyond Moore's law era.  While reading a bit of information stored in the magnetostrictive nanomagnets using a magnetic tunnel junction (MTJ), a material selection issue crops up since magnetostrictive materials in general cannot be utilized as the free layer of the MTJ. This is an important issue since we need to achieve a high magnetoresistance for technological applications. We show here that magnetically coupling the magnetostrictive nanomagnet and the free layer e.g., utilizing the magnetic dipole coupling between them can circumvent this issue. By solving stochastic Landau-Lifshitz-Gilbert equation of magnetization dynamics in the presence of room-temperature thermal fluctuations, we show that such design can eventually lead to a superior energy-delay product.
\end{abstract}

\begin{document}

\flushbottom
\maketitle
\thispagestyle{empty}

\section*{Introduction}

Electric field-induced magnetization switching in strain-coupled multiferroic composites is a promising mechanism that can possibly harness an energy-efficient binary switch replacing the charge-based traditional transistors for our future information processing paradigm~\cite{roy13_spin,roy11}. A voltage applied across such devices strains the piezoelectric layer and the generated stress on the magnetostrictive layer induces a magnetic anisotropy in it~\cite{RefWorks:557,RefWorks:157,RefWorks:558,Refworks:164,Refworks:165,RefWorks:519,roy10,RefWorks:848,RefWorks:865,RefWorks:869,RefWorks:871} and can switch its magnetization~\cite{roy13_spin,roy11,roy11_6,roy13_2}. These straintronic devices operate at room-temperature and the study estimates very promising performance metrics, e.g., energy dissipation of $\sim$1 attojoule (aJ) and sub-nanosecond switching delay, suitable for technological application purposes~\cite{roy11_6,roy11_2}. Experimental efforts to investigate such device functionality has demonstrated the induced stress anisotropy in magnetostrictive nanomagnets~\cite{RefWorks:559,RefWorks:806,RefWorks:836,RefWorks:609,RefWorks:790,RefWorks:868,RefWorks:838}, while the direct experimental demonstration of switching speed (rather than ferromagnetic resonance experiments to get the time-scale) and using low-thickness piezoelectric layers while avoiding considerable degradation of the piezoelectric constants [e.g., $<$ 100 nm of lead magnesium niobate-lead titanate (PMN-PT)]~\cite{RefWorks:823,RefWorks:820} are still under investigation.

\begin{figure}
\centering
\includegraphics{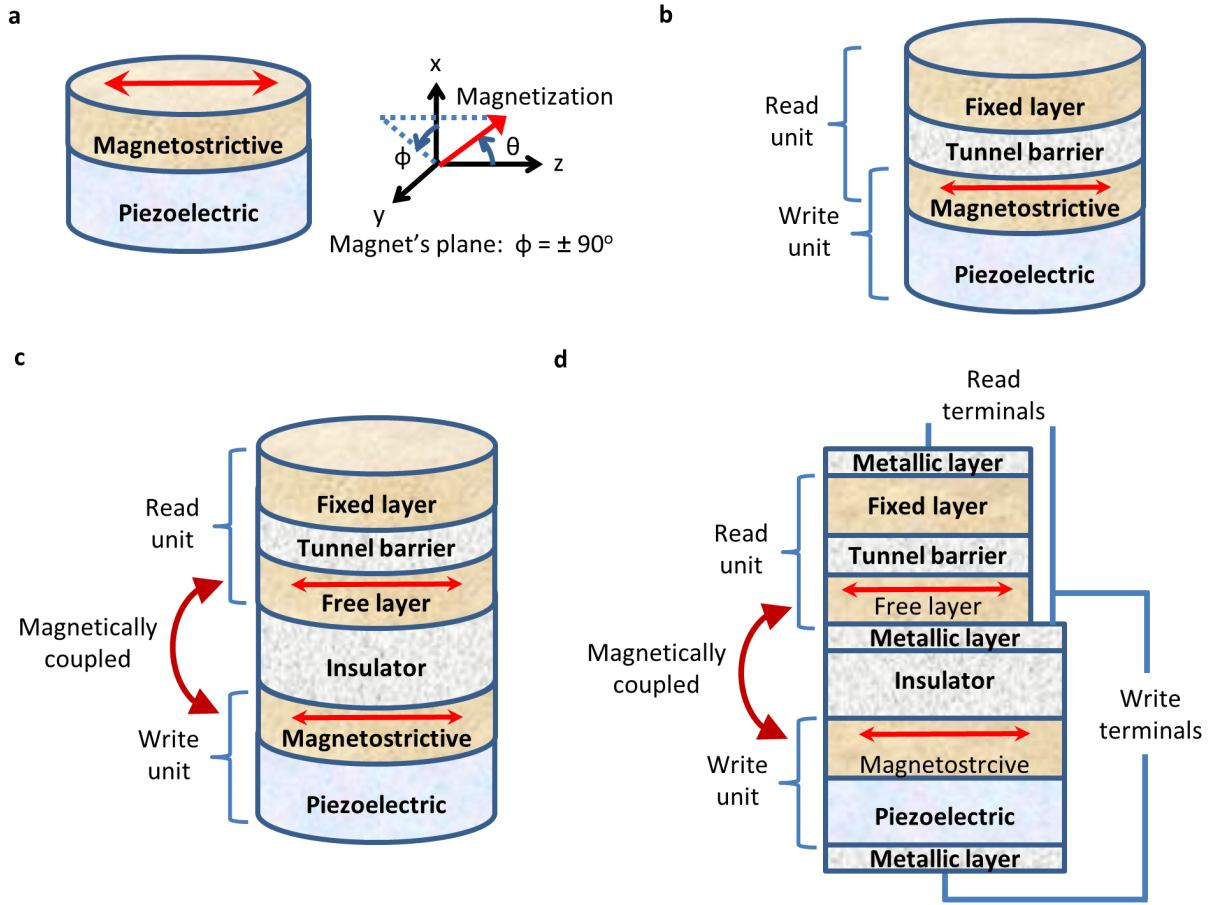}
\caption{\label{fig:read_write_separate} \textbf{Separating read and write units in multiferroic composite devices.} (a) A write unit utilizing multiferroic composites, i.e., piezoelectric-magnetostrictive heterostructures, and axis assignment using standard spherical coordinate system. 
(b) A read unit in the form of a magnetic tunnel junction (MTJ) is incorporated to read the magnetization state of the magnetostrictive nanomagnet, which acts as the free layer in the MTJ. However, materials like half-metals, CoFeB are suitable to constitute the free layer since that leads to a high tunneling magnetoresistance (TMR). Hence, it necessitates to decouple the read unit and write unit in a multiferroic device.
(c) The read unit and the write unit are separated by an insulator but the magnetostrictive nanomagnet and the free layer are magnetically coupled through dipole coupling. Note that the read current flows through the read unit (MTJ) only and the detailed contacts are not shown in this schematic diagram.
(d) A detailed design to make contacts for the read and write units in the proposed design. The read current must flow through the read unit (MTJ) only so that the TMR of the MTJ does not get affected. A metallic layer on the insulator layer can be incorporated to form an equipotential surface, which makes a contact for the free layer in the MTJ. The write terminals are also shown here. Since the piezoelectric layer is much more resistive than that of the insulator, having the insulator in the path does not affect the operation and hence it does not necessitate another metallic layer below the insulator layer for the corresponding write unit terminal.
}
\end{figure}

There are proposals on devising both memory~\cite{roy11,roy11_6,roy13_2} and logic devices~\cite{roy13_spin,roy13x,roy14x} using strain-mediated multiferroic composites by energy-efficient \emph{writing} of a bit of information in the magnetostrictive nanomagnets~\cite{roy14_4}. However, while electrically \emph{reading} the magnetization state of the magnetostrictive nanomagnet using a magnetic tunnel junction (MTJ)~\cite{RefWorks:577,RefWorks:555,RefWorks:572,RefWorks:76,RefWorks:74,RefWorks:33,RefWorks:300}, we need to tackle a material selection issue since the magnetostrictive materials in general cannot constitute the free layer of an MTJ. The widely-used material that is used for the free layer of an MTJ is CoFeB~\cite{RefWorks:786}, which leads to high tunneling magnetoresistance (TMR) of 300\%~\cite{RefWorks:33}. The incorporation of half-metals as the free layer can lead to even better TMR of more than 1000\%~\cite{RefWorks:581}. To tackle this material selection issue, we propose to magnetically couple the magnetostrictive nanomagnet and the free layer of an MTJ, e.g., to utilize the magnetic dipole coupling in between them separated by an insulator (see Fig.~\ref{fig:read_write_separate}). During write operation, as the magnetization of the magnetostrictive layer rotates upon application of stress, the free layer's magnetization also rotates concomitantly and it can be read by an MTJ. Similar methodology of incorporating an insulator for utilizing magnetic dipole coupling has been proposed in the context of input-output isolation for logic design purposes~\cite{RefWorks:754}. Note that the input-output isolation is inherent in multiferroic devices due to the presence of the insulating piezoelectric layer~\cite{roy13x}. We study the effect of this dipole coupling by solving stochastic Landau-Lifshitz-Gilbert equation of magnetization dynamics in the presence of room-temperature thermal fluctuations. The results reveal that such dipole-coupled design can lead to lowering the energy dissipation and a superior energy-delay product. 

\section*{Model}

The nanomagnets are modeled in the shape of elliptical cylinders with the cross-sections lying on the $y$-$z$ plane; the major axis points along the $z$-direction, and the minor axis along the $y$-direction (see Fig.~\ref{fig:read_write_separate}a). In standard spherical coordinate system, $\theta$ is the polar angle and $\phi$ is the azimuthal angle. Any deflection out of magnet's plane ($\phi=\pm90^\circ$) is termed as out-of-plane excursion. We solve the magnetization dynamics using stochastic Landau-Lifshitz-Gilbert (LLG) equation in the presence of room-temperature (300 K) thermal fluctuations. Note that there are two nanomagnets (magnetostrictive and free layer) having a dipole coupling between them. Stress is generated only on the magnetostrictive nanomagnet, hence we have additional stress anisotropy to consider for the magnetostrictive nanomagnet. We will use the subscripts $m$ and $f$ to denote any parameter for the \emph{magnetostrictive} nanomagnet and the \emph{free layer} nanomagnet, respectively.

The magnetization $\mathbf{M_m}$ ($\mathbf{M_f}$) of the magnetostrictive (free layer) nanomagnet has a constant magnitude  but a variable direction, so that we can represent it by a vector of unit norm $\mathbf{m_{m}} =\mathbf{M_m}/|\mathbf{M_m}| = \mathbf{\hat{e}_{r}}$ ($\mathbf{m_{f}} =\mathbf{M_f}/|\mathbf{M_f}| = \mathbf{\hat{e}_{r}}$) where $\mathbf{\hat{e}_{r}}$ is the unit vector in the radial direction in spherical coordinate system represented by ($r$,$\theta$,$\phi$). The other two unit vectors corresponding to the polar angle $\theta$ and the azimuthal angle $\phi$ are $\mathbf{\hat{e}_{\theta}}$ and $\mathbf{\hat{e}_{\phi}}$, respectively.

The potential energies of the \emph{magnetostrictive} nanomagnet and the \emph{free layer} nanomagnet can be expressed, respectively, as
\begin{equation}
E_{total,m} = E_{shape,m} + E_{stress} + E_{dipole} = B_m(\phi_m) \, sin^2 \theta_m + E_{dipole},
\label{eq:total_anisotropy_magnetostrictive}
\end{equation}
and
\begin{equation}
E_{total,f} = E_{shape,f} + E_{dipole} = B_f(\phi_f) \, sin^2 \theta_f + E_{dipole},
\label{eq:total_anisotropy_free_layer}
\end{equation}
where
\begin{subequations}
\begin{align}
B_m(\phi_m) &= B_{shape,m}(\phi_m) + B_{stress}, \\
B_{shape,m}(\phi_m) &= (\mu_0/2) M_{s,m}^2 \Omega_m [(N_{d-yy,m}-N_{d-zz,m}) + (N_{d-xx,m}-N_{d-yy,m})\,cos^2\phi_m],\\ 
B_{stress} 	&= (3/2) \lambda_s \sigma \Omega_m, 
\end{align}
\label{eq:B_shape_stress_magnetostrictive}
\end{subequations}
\begin{subequations}
\begin{align}
B_f(\phi_f) &= B_{shape,f}(\phi_f),\\
B_{shape,f}(\phi_f) &= (\mu_0/2) M_{s,f}^2 \Omega_f [(N_{d-yy,f}-N_{d-zz,f}) + (N_{d-xx,f}-N_{d-yy,f})\,cos^2\phi_f],
\label{eq:shape_free_layer}
\end{align}
\end{subequations}
\begin{equation}
	E_{dipole}(\theta_m,\theta_f,\phi_m,\phi_f) = \cfrac{\mu_0}{4\pi R^3} \, M_{s,m} \Omega_m M_{s,f} \Omega_f \lbrack cos\theta_m cos\theta_f 
	+ sin \theta_m sin \theta_f (sin \phi_m sin\phi_f - 2 cos\phi_m cos\phi_f) \rbrack \displaybreak[3],
\label{eq:dipole}	
\end{equation}
[The dipole coupling between two magnetic moments $\mathbf{M_m}$ and $\mathbf{M_f}$ separated by a distance vector $\mathbf{R}$ can be expressed as~\cite{RefWorks:157} $E_{dipole} = (1/4\pi \mu_0 R^3) \lbrack (\mathbf{M_m.M_f}) - (3/R^2) (\mathbf{M_m.R}) (\mathbf{M_f.R})\rbrack$, where putting $|\mathbf{M_m}|=\mu_0 M_{s,m} \Omega_m$, $|\mathbf{M_f}|=\mu_0 M_{s,f} \Omega_f$, and $\mathbf{R}=R\,{\mathbf{\hat{e}_{x}}}$, we get the equation~\eqref{eq:dipole}.]

\noindent
$M_{s,m}$ ($M_{s,f}$) is the saturation magnetization of the magnetostrictive (free layer) nanomagnet, $\Omega_m$ ($\Omega_f$) is the volume of the magnetostrictive (free layer) nanomagnet, $N_{d-pp,m}$ ($N_{d-pp,f}$) is the component of the demagnetization factor for the magnetostrictive (free layer) nanomagnet along $p$-direction, which depends on the nanomagnet's dimensions~\cite{RefWorks:157,RefWorks:402}, $(3/2)\lambda_s$ is the magnetostrictive coefficient of the single-domain magnetostrictive nanomagnet~\cite{RefWorks:157}, $\sigma$ is the stress on the magnetostrictive nanomagnet, and $R$ is the center-to-center distance between the nanomagnets.
 
The initial orientation of the magnetizations is antiparallel due to dipole coupling between the nanomagnets. From equation~\eqref{eq:B_shape_stress_magnetostrictive}, note that when we apply a sufficient stress (\emph{compressive} stress for materials with \emph{positive} $\lambda_s$ or vice-versa so that the product $\lambda_s\,\sigma$ is \emph{negative}) on the magnetostrictive nanomagnet, the induced stress anisotropy can beat the shape anisotropy of the nanomagnet and rotate its magnetization toward the hard axis~\cite{roy11,roy11_6,roy13_2}. As the magnetization of the magnetostrictive nanomagnet rotates due to the stress anisotropy induced in it, the magnetization of the free layer nanomagnet does also rotate due to the magnetic dipole coupling between the nanomagnets. The magnetizations keep antiparallel orientation when they reach the hard axis. Upon removal of stress from the magnetostrictive nanomagnet, it switches to the opposite direction due to out-of-plane excursion of magnetization~\cite{roy11,roy13_2}. The dipole coupling switches the magnetization of the free layer concomitantly. 

The effective field and torque acting on the magnetostrictive nanomagnet due to the gradient of potential landscape as given by the equation~\eqref{eq:total_anisotropy_magnetostrictive} can be expressed as $\mathbf{H_{eff,m}}  = - \nabla E_{total,m} = - (\partial E_{total,m}/\partial \theta_m)\,\mathbf{\hat{e}_{\theta}} - (1/sin\theta_m)\,(\partial E_{total,m}/\partial \phi_m)\,\mathbf{\hat{e}_\phi}$ and $\mathbf{T_{E,m}} = \mathbf{m_{m}} \times \mathbf{H_{eff,m}}$, respectively. Similarly, the effective field and torque acting on the free layer nanomagnet due to the gradient of potential landscape as given by the equation~\eqref{eq:total_anisotropy_free_layer} can be expressed as $\mathbf{H_{eff,f}} = - \nabla E_{total,f} = - (\partial E_{total,f}/\partial \theta_f)\,\mathbf{\hat{e}_{\theta}} - (1/sin\theta_f)\,(\partial E_{total,f}/\partial \phi_f)\,\mathbf{\hat{e}_\phi}$ and $\mathbf{T_{E,f}} = \mathbf{m_{f}} \times \mathbf{H_{eff,f}}$, respectively.

The thermal field and the corresponding torque acting on the magnetostrictive (free layer) nanomagnet can be written~\cite{RefWorks:186,roy11_6} as $\mathbf{H_{TH,m(f)}} =P_{\theta_{m(f)}}\,\mathbf{\hat{e}_\theta}+P_{\phi_{m(f)}}\,\mathbf{\hat{e}_\phi}$ and $\mathbf{T_{TH,m(f)}}=\mathbf{m_{m(f)}} \times \mathbf{H_{TH,m(f)}}$, respectively, where
\begin{subequations}
\begin{align}
P_{\theta_{m(f)}} &= M_{V,m(f)} [ h_{x,m(f)}\,cos\theta_{m(f)}\,cos\phi_{m(f)} + h_{y,{m(f)}}\,cos\theta_{m(f)} sin\phi_{m(f)}  - h_{z,{m(f)}}\,sin\theta_{m(f)}], \displaybreak[3]\\
P_{\phi_{m(f)}} &= M_{V,{m(f)}} [h_{y,{m(f)}}\,cos\phi_{m(f)} -h_{x,{m(f)}}\,sin\phi_{m(f)}], \displaybreak[3]\\
h_{i,{m(f)}} &= \sqrt{\frac{2 \alpha_{m(f)} kT}{|\gamma| M_{V,{m(f)}} \Delta t}} \;G_{m(f)}{(0,1)} \;\;(i=x,y,z),
\end{align}
\end{subequations}
$\alpha_{m(f)}$ is the phenomenological damping parameter of the magnetostrictive (free layer) material, $\gamma$ is the gyromagnetic ratio for electrons, $M_{V,{m(f)}}= \mu_0 M_{s,{m(f)}} \Omega_{m(f)}$, $\Delta t$ is the simulation time-step, $G_{m(f)}{(0,1)}$ is a Gaussian distribution with zero mean and unit variance for the magnetostrictive (free layer) nanomagnet~\cite{RefWorks:388}, $k$ is the Boltzmann constant, and $T$ is temperature.

The magnetization dynamics of the magnetostrictive (free layer) nanomagnet under the action of the two torques $\mathbf{T_{E,{m(f)}}}$ and $\mathbf{T_{TH,{m(f)}}}$ is described by the stochastic Landau-Lifshitz-Gilbert (LLG) equation~\cite{RefWorks:162,RefWorks:161,RefWorks:186} as 
\begin{equation}
\cfrac{d\mathbf{m_{{m(f)}}}}{dt} - \alpha_{m(f)} \left(\mathbf{m_{{m(f)}}} \times \cfrac{d\mathbf{m_{{m(f)}}}}{dt} \right)
 = -\cfrac{|\gamma|}{M_{V,{m(f)}}} \left\lbrack \mathbf{T_{E,{m(f)}}} +  \mathbf{T_{TH,{m(f)}}}\right\rbrack.
\end{equation}

After solving the above LLG equation, we get the following coupled equations for the dynamics of $\theta_{m(f)}$ and $\phi_{m(f)}$:
\begin{multline}
\left(1+\alpha_{m(f)}^2 \right) \cfrac{d\theta_{m(f)}}{dt} = \cfrac{|\gamma|}{M_{V,{m(f)}}} [ B_{shape,\phi_{m(f)}}(\phi_{m(f)})sin\theta_{m(f)}  - 2\alpha_{m(f)} B_{m(f)}(\phi_{m(f)}) sin\theta_{m(f)} cos\theta_{m(f)} \\
 - T_{dipole,\theta_{m(f)}} - \alpha_{m(f)} T_{dipole,\phi_{m(f)}} 
 + (\alpha_{m(f)} P_{\theta_{m(f)}} + P_{\phi_{m(f)}})],
 \label{eq:theta_dynamics_magnetostrictive}
\end{multline}
\begin{multline}
\left(1+\alpha_{m(f)}^2 \right) \cfrac{d\phi_{m(f)}}{dt} = \cfrac{|\gamma|}{M_{V,{m(f)}}} \cfrac{1}{sin\theta_{m(f)}} [\alpha B_{shape,\phi_{m(f)}}(\phi_{m(f)})sin\theta_{m(f)} + 2 B_{m(f)}(\phi_{m(f)}) sin\theta_{m(f)} cos\theta_{m(f)} \\
   + \alpha_{m(f)} T_{dipole,\theta_{m(f)}} + T_{dipole,\phi_{m(f)}} 
	 - \{sin\theta_{m(f)}\}^{-1} (P_{\theta_{m(f)}} - \alpha_{m(f)} P_{\phi_{m(f)}})] \qquad (sin\theta_{m(f)} \neq 0),
  \label{eq:phi_dynamics_magnetostrictive}
\end{multline}
where
\begin{subequations}
\begin{align}
B_{shape,\phi_{m(f)}}(\phi_{m(f)}) &= -\cfrac{\partial B_{shape,{m(f)}}(\phi_{m(f)})}{\partial \phi_{m(f)}} = (\mu_0/2) \, M_{s,{m(f)}}^2 \Omega_{m(f)} (N_{d-xx,{m(f)}}-N_{d-yy,{m(f)}})sin(2\phi_{m(f)}),\\
T_{dipole,\theta_{m(f)}} &= \cfrac{1}{sin \theta_{m(f)}}\,\cfrac{\partial E_{dipole}}{\partial \phi_{m(f)}},\\
T_{dipole,\phi_{m(f)}} &= \cfrac{\partial E_{dipole}}{\partial \theta_{m(f)}}. 
\end{align}
\end{subequations}

The magnetization dynamics of the two nanomagnets represented by the equations~\eqref{eq:theta_dynamics_magnetostrictive} and~\eqref{eq:phi_dynamics_magnetostrictive} are coupled through the dipole coupling [see equation~\eqref{eq:dipole}]. These coupled equations are solved numerically to track the trajectories of the two magnetizations over time.

The internal energy dissipation in the magnetostrictive (free layer) nanomagnet due to Gilbert damping can be expressed as $E_{d,{m(f)}}=\int_0^\tau P_{d,{m(f)}}(t)\,dt$, where $\tau$ is the switching delay and the instantaneous power dissipation can be calculated as
\begin{equation}
P_{d,{m(f)}}(t) = \cfrac{\alpha_{m(f)} \, |\gamma|}{(1+\alpha_{m(f)}^2) M_{V,{m(f)}}} \, \left| \mathbf{T_{E,{m(f)}}}(t)\right|^2.
\label{eq:power_dissipation_magnetostrictive}
\end{equation}
We sum up these two internal energy dissipations $E_{d,m}$ and $E_{d,f}$ alongwith the energy dissipation due to applying voltage (which is miniscule~\cite{roy11_2,roy11_6}) to determine the total energy dissipation.

\begin{figure}
\centering
\includegraphics{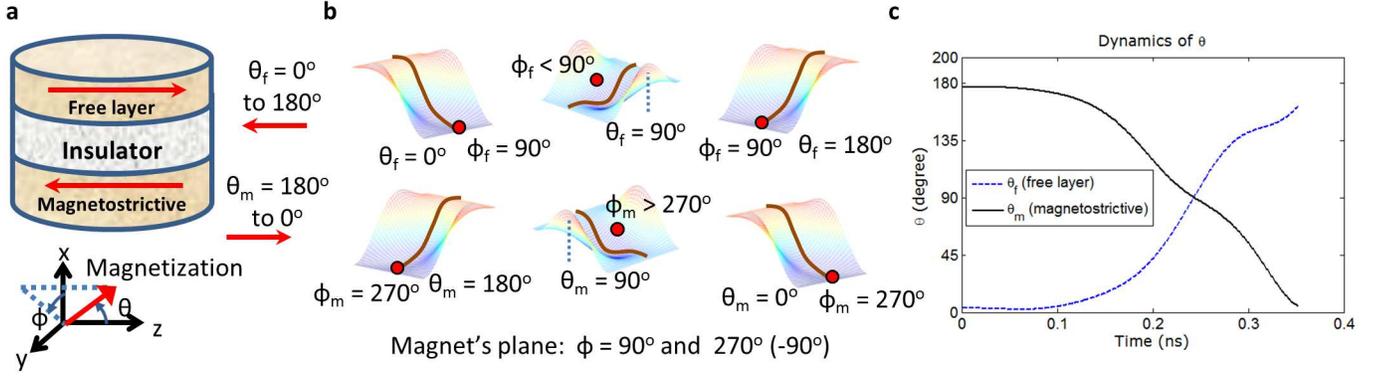}
\caption{\label{fig:switching_180_deg} 
\textbf{Magnetization switching in the magnetostrictive and free layer nanomagnets upon application of stress on the magnetostrictive nanomagnet.}
(a) Magnetizations of the magnetostrictive and free layer nanomagnets are antiferromagnetically coupled due to dipole coupling.
(b) Potential landscapes (bold lines correspond to magnets' planes) and positions of the magnetizations for both the nanomagnets during switching. When no stress is applied, the potential landscapes of the nanomagnets are monostable due to dipole coupling and the magnetizations are antiferromagnetically coupled. If a sufficient stress is exerted on the magnetostrictive nanomagnet, the magnetizations come to their hard-axes and still remain antiferromagnetically coupled, however, they are deflected out-of-plane. This out-of-plane excursion eventually leads to the full 180$^\circ$ switching when the stress is released/reversed. 
(c) The LLG simulation results show that both the magnetizations successfully switch by 180$^\circ$. A stress of 30 MPa is applied on the magnetostrictive nanomagnet and the stress is reversed to aid the switching speed when the magnetization comes to the hard axis. The ramp (rise and fall) time of stress is 60 ps. No thermal fluctuations is considered but the deflection in initial orientations of the magnetizations is taken as $\sim$3.25$^\circ$, which is the thermally mean value due to room-temperature (300 K) thermal fluctuations when no stress is active. When $\theta_m$ becomes $\leq 5^\circ$, the switching is deemed to have completed and the switching delay is recorded as 0.351 ns.
}
\end{figure}

\section*{Results}

The magnetostrictive nanomagnet is made of polycrystalline Terfenol-D and it has the following material properties -- saturation magnetization ($M_{s,m}$):  8$\times$10$^5$ A/m, Gilbert damping parameter ($\alpha_m$): 0.1, Young's modulus (Y): 80 GPa, magnetostrictive coefficient ($(3/2)\lambda_s$): +90$\times$10$^{-5}$ (Refs.~\citenum{RefWorks:179,RefWorks:176,RefWorks:178,roy11_6}), and Poisson's ratio ($\nu$): 0.3 (Ref.~\citenum{RefWorks:821}). The free layer nanomagnet is made of widely-used CoFeB, which has the following material properties -- Gilbert damping parameter ($\alpha_f$): 0.01, saturation magnetization ($M_{s,f}$):  8$\times$10$^5$ A/m~\cite{RefWorks:413}. The dimensions of both the single-domain nanomagnets are chosen as $100\,nm \times 90\,nm \times 6\,nm$~\cite{RefWorks:402,RefWorks:133}, and the center-to-center distance between the nanomagnets is $R=40\,nm$. For the piezoelectric layer, we use PMN-PT, which has a dielectric constant of 1000, $d_{31}$=--3000 pm/V, and $d_{32}$=1000 pm/V~\cite{RefWorks:790}. We assume the piezoelectric layer's thickness $t_{piezo}$=24 nm (Ref.~\citenum{roy11_6}) and thus $V=1.9$ mVs (2.9 mVs) of voltages would generate 20 MPa (30 MPa) compressive stress [$\sigma=Y\,d_{eff}\,(V/t_{piezo})$, where $d_{eff}=(d_{31}-d_{32})/(1+\nu)$] in the magnetostrictive Terfenol-D layer. Note that avoiding considerable degradation of the piezoelectric constants at such low-thickness (24 nm) piezoelectric layers is under research~\cite{RefWorks:823,RefWorks:820}. Modeling the piezoelectric layer as a parallel plate capacitor, the capacitance C=2.6 fF and thus $CV^2$ energy dissipation turns out to be $<$ 0.1 aJ. This is the basis of ultra-low-energy computing using these multiferroic devices~\cite{roy11,roy13_spin,roy11_6,roy13x,roy14x}.

When the magnetizations of the magnetostrictive and free layer nanomagnets are \emph{exactly} aligned to their easy axes (e.g., $\theta_m=180^\circ$ and $\theta_f=0^\circ$), the torques acting on the magnetizations are \emph{exactly} zero and hence only thermal fluctuations can deflect the magnetizations from their initial orientations. When no stress is active on the magnetostrictive layer, we solve the stochastic LLG equation in the presence of room-temperature thermal fluctuations to determine the distributions of the magnetizations' initial orientations and calculate the mean orientations of the magnetizations ($\sim$3.25$^\circ$). The initial distribution of magnetization is a Boltzmann distribution and matching the numerically calculated mean orientation of magnetization with the one calculated from the equipartition theorem depicts the validity of incorporating thermal fluctuations~\cite{roy13_2}.

Figure~\ref{fig:switching_180_deg} shows that the magnetizations of the magnetostrictive nanomagnet and the free layer nanomagnet rotate concomitantly upon application of 30 MPa stress on the magnetostrictive nanomagnet. The magnetizations of two nanomagnets come to their respective hard axes and remain antiferromagnetically coupled. The initial values of the azimuthal angles $\phi_{m,init}$ and  $\phi_{f,init}$ are chosen as 270$^\circ$ and 90$^\circ$, respectively, however, they can be just opposite too, which is equally possible. During the course of magnetization dynamics, the exerted stress rotates the magnetization of the magnetostrictive nanomagnet out-of-plane and subsequently the magnetization of the free layer also gets deflected out-of-plane due to dipole coupling, as depicted in the Fig.~\ref{fig:switching_180_deg}b. This out-of-plane excursion increases the switching speed tremendously and creates an intrinsic \emph{asymmetry} to facilitate a complete 180$^\circ$ switching of the magnetizations \emph{deterministically} even in the presence of thermal fluctuations~\cite{roy13_2}. The LLG simulation results as shown in the Fig.~\ref{fig:switching_180_deg}c depict that both the magnetizations have completed full 180$^\circ$ switching.

The magnetization switching procedure described above requires to \emph{read} the magnetization state using MTJ to sense when magnetization reaches \emph{around} $\theta_m=90^\circ$ (since room-temperature thermal fluctuations make the traversal time a wide distribution), so that stress can be brought down thereafter~\cite{roy13_2}. Note that there is tolerance around $\theta_m=90^\circ$, i.e., stress does not need to be withdrawn \emph{exactly} at $\theta_m=90^\circ$ since it is shown that the internal magnetization dynamics provides such tolerance~\cite{roy13_2}. These are purely dynamical phenomena contrary to steady-state analysis. Any additional element for comparison can be built using these energy-efficient multiferroic devices in general~\cite{roy_analog_conf}. Note that researchers are trying to replace the traditional \emph{switch} based on charge-based transistors by a new possible ``ultra-low-energy'' \emph{switch} (e.g., using multiferroic composites). Therefore, any circuitry can be built with the energy-efficient switch itself rather than the conventional transistors. Usually, it requires several peripheral circuitry in conjunction with the basic switch in a system~\cite{rabae03,pedra02}.  While researchers report on the performance metrics of the basic switch itself, the total energy dissipation considering the other required circuitry does not change the order of energy dissipation, utilizing the respective devices~\cite{rabae03,pedra02}. This was the understanding while claiming energy-efficiency using such magnetization switching mechanism~\cite{roy13_spin,roy11,roy11_2,roy13_2,roy11_6} and computing methodologies~\cite{roy13_spin,roy13x,roy14x} based on such switching methodology. 

It may be possible to harness more asymmetry in the system apart from the intrinsic asymmetry due to out-of-plane excursion as described above so that the sensing mechanism for dynamic withdrawal of input voltage may not be necessary. Interface and exchange coupling can also provide asymmetry during switching~\cite{roy14_2}, particularly it helps to maintain the direction of switching rather than toggling~\cite{roy13_2} the magnetization direction, and it does not require any sensing procedure~\cite{roy14_2}. 

\begin{table*}[t]
\small
\begin{tabular}{l c c c c c c c c c c}
Case & Free layer? & Stress (MPa) & $\tau_{mean}$ (ns) & $\tau_{std}$ (ns) & $E_{m}$ (aJ) & $E_{f}$ (aJ) & $E_{d}$ (aJ) & $`CV^2\textrm'$ (aJ) & $E_{total}$ (aJ) & $E_{total}\,\tau_{mean}$ (aJ-ns)\\
\hline
(a) & No   & 20 & 0.444 & 0.080 & 0.89 & NA   & 0.89 & 0.0297 & 0.9296 & 0.4127\\
(b) & Yes  & 20 & 0.529 & 0.120 & 0.62 & 0.07 & 0.69 & 0.0297 & 0.7296 & 0.3860\\
(c) & Yes  & 30 & 0.379 & 0.080	& 0.81 & 0.08 & 0.89 & 0.0669 & 0.9792 & 0.3711\\
(d) & No   & 30 & 0.368 & 0.064	& 1.09 & NA   & 1.09 & 0.0669 & 1.1792 & 0.4340\\
\end{tabular}
\caption{\label{tab:comparison} Performance metrics for four different cases considered. Cases (a) and (d) [corresponding to 20 MPa and 30 MPa stress, respectively] do not consider the additional free layer, while the cases (b) and (c) [corresponding to 20 MPa and 30 MPa stress, respectively] consider the free layer magnetically coupled to the magnetostrictive nanomagnet. With the introduction of the free layer at the same stress level, the switching delay metrics (mean and standard deviation) get worse while it dissipates less energy and leads to less $E_{total}\,\tau_{mean}$ product. Case (c) has the lowest $E_{total}\,\tau_{mean}$.}
\end{table*}

To understand the effect of incorporating the dipole-coupled free layer on the performance metrics, we solve the stochastic LLG equation~\cite{RefWorks:186} at room-temperature (300 K) and tabulate the performance metrics in Table~\ref{tab:comparison} for four different cases. For each case, we perform a moderately large number (10,000) of simulations and when the magnetization of the magnetostrictive nanomagnet reaches $\theta_m \leq 5^\circ$, the switching is deemed to have completed and the switching delay $\tau$ for that trajectory is recorded. Then we determine the following performance metrics: mean value of switching delay ($\tau_{mean}$), standard deviation of switching delay ($\tau_{std}$), the mean values of the energy dissipations $E_{m}$ (in the magnetostrictive layer) and $E_{f}$ (in the free layer) due to Gilbert damping in the magnets, the energy dissipation due to applying voltage $`CV^2\textrm'=3CV^2$ (since stress is reversed~\cite{roy11_6,roy13_2}), and the total energy dissipation $E_{total}=E_d + `CV^2\textrm'$, where $E_d=E_m+E_f$.

For the case (a), we do not have any additional free layer and just consider the switching in the magnetostrictive nanomagnet, while for the case (b), we do have the free layer. For both the cases (a) and (b), the stress is 20 MPa. We also consider the distribution of initial orientation for the case (a) and the mean value of deflection of initial orientation from the easy axis turns out to be $\sim$3.9$^\circ$. This value is higher than that of the case (b) [$\sim$3.25$^\circ$], when the dipole-coupled free-layer is introduced. The reason behind is that the dipole coupling energy confines both the magnetizations more in their respective potential wells so the magnetizations' deflection is less. Hence, while considering switching \emph{with dipole coupling}, magnetizations on average start nearer from the easy axis and therefore it takes more time for switching to be completed~\cite{roy11_6}. This is reflected in the mean value of switching delay if we compare it for the cases (a) and (b). By fixing the same initial orientation of magnetization for the cases (a) and (b), it is noticed that dipole coupling in fact speeds up the switching process so the increase in switching delay for the case (b) is \emph{entirely} due to the less deflection in the initial orientation of magnetization as described above. The case (b) incurs less energy dissipation in total, which indicates the delay-energy trade-off, i.e., slower switching dissipates less energy. Note that case (b) has lower energy-delay product ($E_{total}\,\tau_{mean}$) compared to the case (a).

\begin{figure}
\centering
\includegraphics{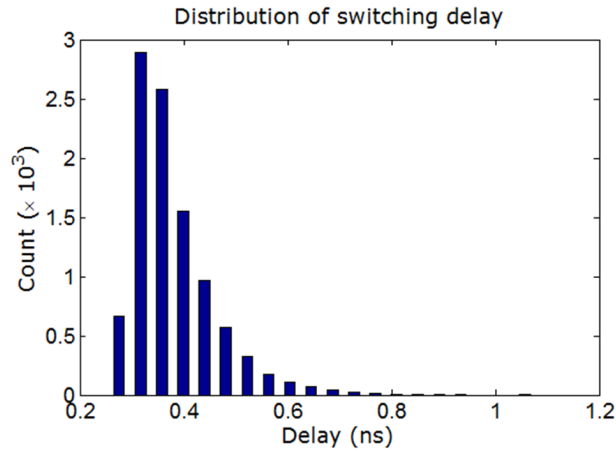}
\caption{\label{fig:distribution_180_deg_30MPa} 
\textbf{An illustrative distribution of switching delay for the case of magnetostrictive nanomagnet.} A moderately large number (10,000) of simulations performed in the presence of room-temperature (300 K) thermal fluctuations. The dipole-coupled free layer is considered in this case and 30 MPa stress is applied with 60 ps ramp (rise and fall) time. This wide distribution is caused by the following two reasons: (1) thermal fluctuations make the initial orientation of magnetization a distribution, and (2) thermal kicks during the switching makes the switching delay a distribution too; the former one has a higher effect than that of the latter. The mean and standard deviation of this distribution are 0.379 ns and 0.080 ns, respectively.
} 
\end{figure}

To investigate the performance metrics with the incorporation of the dipole-coupled free layer further, we increase the stress to 30 MPa and tabulate the results as case (c) in the Table~\ref{tab:comparison}. We note that the mean switching delay has got reduced compared to the case (a) while they incur the same amount of energy dissipation due to Gilbert damping. We plot the corresponding switching delay distribution for the case (c) in the Fig.~\ref{fig:distribution_180_deg_30MPa}. Such distribution can be achieved experimentally by time-resolved measurements~\cite{RefWorks:837}. It needs to be pointed out that we can generate a maximum amount of stress on the magnetostrictive layer dictated by the maximum strain induced in it, so we also consider 30 MPa stress without the free layer and tabulate the results as case (d). The mean switching delay $\tau_{mean}$ for case (d) is very close to that of case (c), but it has the highest energy dissipation $E_{total}$ (and also $E_{total} \tau_{mean}$) among the four cases considered, while case (c) has the lowest $E_{total} \tau_{mean}$. Assuming a performance metric $E_{total}\,\tau$, where $\tau=\tau_{mean}+10\,\tau_{std}$, the case (c) still would have the lower product compared to the case (d). 

Note that the switching delay is optimized for lowest value with respect to dipole coupling strength, which can be tuned by varying the thickness of the insulator separating the magnetostrictive nanomagnet and the free layer. With higher thicknesses of the insulator, the magnetizations of the two layers do not quite rotate concomitantly and thus the switching delay is increased, while for lower thicknesses, higher dipole coupling rotates the magnetization  out-of-plane  so much that it leads to precessional motion and it increases the net switching delay. 

\section*{Discussion}

We have addressed the material selection issue while reading out the state of the magnetostrictive nanomagnet in a multiferroic composite. The proposed design provides us the flexibility to use the best materials for the magnetostrictive nanomagnet and the free layer in an MTJ separately. Rather than dipole coupling, we can also utilize exchange coupling between the two nanomagnets to magnetically couple them. Note that this is a general strategy, which can be also utilized in spin-transfer-torque switching of nanomagnets where we can use the switching nanomagnet to be made of CoFeB and the free layer to be made of half-metals for higher TMR. Hence, it will motivate experiments and further theoretical studies on this front. Moreover, it turns out that this design also enhances the energy-delay performance metric. Such ultra-low-energy and non-volatile (leading to instant turn-on computer) computing paradigm is particularly promising to become the staple of our future information processing systems.


\section*{Acknowledgements}

This work was supported in part by FAME, one of six centers of STARnet, a Semiconductor Research Corporation program sponsored by MARCO and DARPA. 

\section*{Additional information}

The author declares no competing financial interests. Correspondence and requests for materials should be addressed to K.R. (email: royk@purdue.edu).

\end{document}